\documentclass[ ]
{aipproc}

\layoutstyle{8x11single}

\begin{document}

\title{A Realist Interpretation \newline of the Quantum Measurement Problem}


\author{Xiaolei Zhang}{
address={~~~~~~~~~~~~~~~~~~~~~~~~   U.S Naval Research Laboratory, Remote Sensing Division, 
\newline 4555 Overlook Ave. SW, Washington, DC 20375, USA}
}

\begin{abstract}
A new, realist interpretation of the quantum measurement processes is given.  
In this scenario a quantum measurement is a non-equilibrium phase transition in 
a ``resonant cavity'' formed by the entire physical universe including all 
its material and energy content.  Both the amplitude and the phase of the 
quantum mechanical wavefunction acquire substantial meaning in this picture, 
and the probabilistic element is removed from the foundations of quantum 
mechanics, its apparent presence in the quantum measurement process is viewed 
as a result of the sensitive dependence on initial/boundary conditions of the 
non-equilibrium phase transitions in a many degree-of-freedom system.
The implications of adopting this realist ontology to the clarification 
and resolution of lingering issues in the foundations of quantum mechanics,  
such as wave-particle duality, Heisenberg's uncertainty relation, Schrodinger's
Cat paradox, first and higher order coherence of photons and atoms, virtual 
particles, the existence of commutation relations and quantized behavior, etc., 
are also presented.
\end{abstract}

\maketitle

\section{Introduction}

Since its very inception, quantum mechanics was plagued at its
foundation with the so-called ``interpretation problem''.  Unlike a
classical measurement, the result of a quantum measurement in general
can only be predicted in a probabilistic sense.  The quantum mechanical
wavefunction is the most quantitative result from a quantum calculation,
and one can hope for a definitive answer only if the measurement is
towards an eigenvalue of a system already settled onto its
corresponding eigenstate.  If this is not so, the underlying
system is assumed to subsequently ``collapse'' onto an eigenstate
of the measurement operator, with the probability for the choice of
eigenstate given by the absolute square of the wavefunction projected
onto the the eigenstate being measured.
                                                                                
Debates over the past 80 years over the underlying reality/or the lack
thereof of the wavefunction collapse had resulted in a wide spectrum of
possible explanations, from the earliest orthodox Copenhagen
Interpretation\cite{Bohr1928, Born1926,Heisenberg1927}, 
to the ``Many Worlds Interpretation\cite{Everett1957, DeWitt1970},
and onward to the different versions of the so-called ``Decoherence
Theories''\cite{VonNeumann1932, Wigner1963,Zurek2002} which advocate 
a two-stage wavefunction collapse, with the second stage involving 
the intervention of a conscious observer.  In a previous 
article\cite{Zhang2005}, we had shown that
all of these previous proposals amounted to giving up realism at
certain stages of the interpretation, which often lead to pathological
cases such as the infamous ``Schrodinger's Cat paradox''\cite{Schrodinger1935}.
Other candidate interpretations such as deBroglie and Bohm's pilot wave
theories\cite{deB1960,Bohm1952} are contradicted by the image provided by
modern quantum field theories on the nature of fundamental particles
as non-local field resonances\cite{Cao1998}.

In this article, we argue that the resolution of the quantum
measurement problem lies in its analogy to a class of nonequilibrium
phase transitions in the classical world and the resulting formation
of the so-called ``dissipative structures''\cite{Prigogine1977,Zhang1998,
Zhang2005}.

\section{The New Ontological View}

The universe is an open and evolving system.  If we insist on a
realist interpretation of the quantum processes, and insist
also a classical-type ontology for quantum phenomena which connects with
our intuition, we need to seek the origin of the quantized behavior
in some forms of resonance interaction as well.  Since physical
constants such as Planck's constant and the speed of light are universal, and
different types of interaction processes in vastly different environments
give the same elementary-particle properties, these universal characteristics
are expected to be produced through the actions of the entire matter
and energy content of the universe.  Such a view is closely related to
Mach's principle for explaining the origin of mass and inertia\cite{Mach1893},
and forms the starting point of the new ontological view of
quantum mechanics we are presenting in this paper.
                                                                                
We now summarize the main hypotheses of this work:
\begin{itemize}
\item It is assumed that a generalized ``Mach's Principle''
governs the operation of the physical universe. Features of the
local physical interactions,
including the values of fundamental constants and the forms of
physical laws, are determined by the global distribution
of all the matter and energy content in the universe.
Such a view is shared by many contemporary working
physicists\cite{Barbour1995, Sachs2003}.
\item The quantized nature of fundamental processes originates
from nonequilibrium phase transitions in the universe
resonant cavity.  The usual quantization procedure by enforcing
the commutation relation is equivalent to establishing modal
closure relation in the universe resonant cavity.  Uncertainty relations
are the phenomenological derivatives of the corresponding commutation
relations and thus have no independent fundamental significance.
\item Due to the successive nature of phase transitions
the physical universe is organized into hierarchies in
both its laws and its phenomena.  The hierarchy in the
physical laws is reflected in the spontaneous breaking of
gauge symmetry in the formation of physical laws, as
well as in the division of the domains of validity
of laws governing different branches of physics and chemistry.
Similar hierarchy in the physical phenomena is responsible for forming
the quantum and classical division, as well as the micro- and macroscopic
division.  Macroscopic structures, being stable resonance
features in the large scale, are capable of resisting the
diffusion/smearing tendency of pure quantum states governed by
Schrodinger-type wave equations.
\item A quantum measurement process happens under appropriate
boundary conditions such that an irreversible nonequilibrium
phase transition is induced in the joint system of the object being
measured, the measuring instrument and the rest of the universe.
\item A quantum mechanical wavefunction describes the substantial
distribution of the underlying matter of a specific modal type
in the configuration space.  Its absolute square gives the probability
for obtaining a particular result in the measurement phase transition.
Its phase encodes the influence of the environment including the
different types of force fields, and the gradient of the phase determines the
subsequent evolution of the wavefunction.  The probabilistic element
is thus removed from the ontology of quantum mechanics.
\item In this picture the vacuum fluctuations are the ``residuals''
after forming the ``whole'' numbers of quasi-stationary modes in the open,
nonequilibrium universe cavity.  The fluctuations in the vacuum reflect
the continuous energy exchange with the ``whole'' modes needed to
sustain these modes.
\end{itemize}
                                                                                
\section{Implications on the Foundation of Quantum Mechanics}

\subsection{The Quantum Mechanical Wavefunction}
                                                                                
In the current ontology, the quantum mechanical wavefunction or
the so-called probability wave becomes a physical entity.
The fact that the wavefunction exists in the 3n-dimensional
configuration space does not pose a problem, since this signifies only
the interchange and interrelation of the parts and parcels of the
modal content among a multi-particle quantum state.  
The dispersion of the position-eiegnstate wavepacket\cite{Schrodinger1926}
is here seen as a natural tendency for a localized ``particle''
to evolve towards plane-wave-like momentum eigenstate\cite{Zhang2005}.

Whereas the amplitude of the wavefunction is connect to the modal
density in the configuration space, and thus leads to the
probability of a particular measurement outcome,
the spatial variation of the phase of the wavefunction characterizes
the probability flux, and thus determines its subsequent evolution.
The phase of the quantum mechanical wavefunction also carries imprints of
various kinds interaction fields, and is the medium through which the
force fields apply their influence to the wavefunction.
                                                                                
\subsection{Evolution of the Quantum States}

The wavefunction of a quantum observable usually spreads out
in infinite space, and the interaction between the different
quantum modes are global in nature. Therefore
quantum mechanics is formulated in the Hilbert space which
is a natural domain to describe global modal relations.
                                                                                
Traditionally, a general quantum evolution involves the two-stage
process of a continuous unitary evolution
which is covariant and energy conserving, followed by a discontinuous
``wavefunction collapse'', represented by the reduction of the
wavefunction into an eigenstate of an observable.
It was the second step, i.e., the wavefunction collapse process,
which acquired a more substantial meaning in the new ontology.
                                                                                
Even in the unitary evolution phase, certain features of the global
interaction can already be discerned.
For a state which is not an eigenstate of the energy operator,
expand it in terms of the eigenstates of an observable $\bf{\alpha}$ that
commutes with the Hamiltonian H, we have\cite{Sakurai1985}
\begin{equation}
\Psi(\vec{x}, t) = \int d^3x' K(\vec{x},t, \vec{x'},t_0) \Psi(\vec{x'},t_0)
,
\end{equation}
\label{eq:5}
where
\begin{equation}
K(\vec{x},t; \vec{x'},t_0) = \sum_{{\bf \alpha}} <\vec{x}|{\bf \alpha}>
<{\bf \alpha}|\vec{x'}> \exp
{{ {-i E_{{\bf \alpha}} (t - t_0)} \over \hbar}}
,
\label{eq:6}
\end{equation}
which represents the transition amplitude between the relevant
eigenstates.  We see that as the
unitary evolution proceeds the relative weights of the constituent
energy eigenstates vary with time, i.e., there is a kind of
transition process happening continuously between the constituent eigenstates.
The main difference between the transitions during the unitary evolution,
and the transitions during a true quantum measurement, is whether
the transition is controlled by a continuous and predictable
evolution of the {\em complex probability amplitude}, or else a discontinuous
settling onto a definitive eigenstate of the measurement operator
with certain {\em probability} (equal to square of the probability amplitude).
                                                                                
Even though an eigenvalue of a stationary
state is a constant (call it $\alpha$), the eigenfunction is in general
time dependent (the particular form of the evolution of the
eigenfunction, if it is a simultaneous eigenstate of the Hamiltonian, 
is $|{\bf \alpha}> \exp ({{-i E_{\alpha} t} \over {\hbar}}$)\cite{Sakurai1985}.
Therefore a quantum mechanical stationary state is a kind of dynamical
equilibrium in constant evolution (including a constant exchange with
the quantum vacuum), consistent with the nonequilibrium
stationary state picture we have proposed.
                                                                                
In practice, apart from stationary states and freely-evolving
wavefunctions, the density matrix formalism\cite{VonNeumann1927} 
had also been employed to describe quantum systems that are thought to be 
statistical mixtures.  In the current ontology, no physical system
is actually in a statistical mixture state.  The apparent
success of the density matrix approach is understood as the intrinsic
harmonic nature of the evolution of the parts and parcels of constituent
quasi-stationary states -- thus the effect of time averaging
mimics the effect of ensemble averaging.
                                                                                
\subsection{Extent of Quantum Measurement}
                                                                                
A quantum measurement is in general a non-local process. 
Quantum measurements involve not only the objects being measured,
the measuring apparatus, but also involve
the ``give and take'' with the rest of the universe. 
This ``give and take'' with the rest of the universe accounts for the apparent
violation of energy conservation in many quantum measurement 
processes\cite{Zhang2005} (including the position-momentum measurement pair).
It also provides a natural explanation to the paradoxical fact that the
accelerated electrons radiate in certain cases (as when they travel freely
in straight lines) and not in other cases (as when they circulate
an atomic nuclei in bound states).

A pure quantum mechanical resonance (such as a photon) remains a
modal resonance spread out in space until the moment
of detection.  Detection shows quantum behavior because the
``remainder'' of the stuff in the universe cavity shows quantum behavior.
The detected particle is in general no longer the same particle
during propagation because of the exchange with the universal background.
                                                                                
The strongest support for the involvement of the universe resonant
cavity during the quantum measurement process is actually the consistency
of emerging elementary-particles' characteristics throughout different types
physical processes (such as the constancy of electron charge and mass
whether it is created out of neuton decay, or else out of the
electron/positron pair production from energetic photons).
Without a global resonant cavity to determine
the modal characteristics, we would not have such things as elementary
particles, or fundamental constants themselves.
The identity of particles is the result of their being the same mode,
and the properties of the fundamental particles are created at the
moment of phase transition, since before this transition
an electron, say, does not already exist inside a neutron.
                                                                                
\subsection{Wave-Particle Duality}
                                                                                
The wave-particle duality is manifest most clearly in the de Broglie
relation $ p = h/\lambda$, and Planck relation $ E=h \nu$,
each equation on one side indicates a pure wave characteristic 
($\nu$ and $\lambda$)
and on the other side a pure particle characteristic ($E$ and $p$).
The seeming contradictory characteristics is easily clarified in the
new ontology:  A particle is more of a pure resonance
when it is a wave mode and is is spread out.  When it is a localized particle,
it is in a mixed resonant state, or the superposition of pure states.
                                                                                
In fact, in quantum field theories, only the fields are localized,
but field quanta are spatially extended.  These spatially distributed
field quanta arrive from the first approximation of the solution
of field equations in the non-interacting limit, which simplifies
analysis and is the source of the name ``particle''\cite{Cao1998}.
                                                                                
\subsection{Uncertainty Principle and Commutation Relations}
                                                                                
In general, the uncertainty relations can be shown to originate
from the corresponding equality relations
linking the commutators and anticommutators
of the quantum observables A and B as\cite{Sakurai1985}
\begin{equation}
|< \Delta A \Delta B>|^2 = { 1 \over 4} |<[A, B]>|^2 +
{1 \over 4} |<\{\Delta A, \Delta B\}>|^2
,
\label{eq:11}
\end{equation}
Furthermore, from Schwarz inequality
$
< \Delta A>^2 <\Delta B>^2 ~~ \ge
|< \Delta A \Delta B>|^2
,
$
the usual uncertainty relation can be arrived at:
\begin{equation}
< \Delta A^2> <\Delta B^2>
~~
\ge
{ 1 \over 4} |<[A, B]>|^2
.
\end{equation}
The uncertainty principle is thus just another way of writing
the commutation relations, which themselves are deterministic.
The uncertainty principle itself seemed to later acquire more
prominence in the discussions of quantum phenomena only because of its
intimate relation to the probabilistic outcome of quantum
measurements.
                                                                                
\subsection{Quantum Vacuum}
                                                                                
After quantizing space with a set of modes using the commutation or
anticommutation relations, we expect to end up with some leftovers,
as is typical for the formation of nonequilibrium quasi-stationary
modes\cite{Zhang1998}.  
We proposed that these leftover fractional modal content
in the quantization process is the constituents of vacuum fluctuations.
                                                                                
The vacuum field fluctuates because the resonant components keep evolving
in the nonequilibrium universe cavity, just as in another example of
such a nonequilibrium dissipative structure, that of the spiral structure in
galaxies\cite{Zhang1998}, where the individual star's trajectory keeps evolving
and moving in and out of the spiral pattern even though total energy is
conserved and the spiral density wave is meta-stable.
                                                                                
Many effects which so far have been attributed to the quantum
vacuum, such as the scale dependence in the renormalization group approach,
the Casimir force, the virtual particles and the lifetime of atomic levels, 
as well as the spontaneous emission of radiation, can equally be thought
of as due to the influence of the rest of the matter distribution in the
universe.  For example, the addition of metal plates as in the
Casimir effect changes the boundary
condition of the entire vacuum, force is thus needed to put the plates in.
Virtual particles are those which appear in a quantum electrodynamic
calculation and do not satisfy energy and angular momentum conservation:
They are ``not on the mass shell'' and are represented by the internal lines
in Feynman diagrams.  Their existence is another indication
that a quantum phase transition involves the rest of the universe to
``close the loop'', and the conservation relation is restored for resonant
interactions only when the phase transition is complete (this last condition
is in fact not yet met by the current pertubative QED, in the strong
coupling case or in higher order calculations, likely indicating the
inherent failure for a local field theory to become self-consistent).
                                                                                
This view of the vacuum also provides a possible explanation of why
some of these very same effects (including Lamb shifts,
Casimir effects, spontaneous emission, van der Waals forces, and
the fundamental linewidth of a laser) can be explained equally successfully
by adopting either the vacuum-fluctuation point of view
or the source-field point of view\cite{Milonni1994}.
                                                                                
The relation of field quantization and vacuum fluctuation  may also be
related to the ``fluctuation-dissipation theorem''.  The dissipative
leak into the vacuum is needed to maintain the stability of the
nonequilibrium modes.   The environment needs to be constantly
evolving in order for the fundamental resonances to be stable.
So the un-saturatedness and the constant evolving nature of the universe
maybe a prerequisite for setting up the fundamental laws as we
observe today.
                                                                                
\subsection{First and Higher Order Coherence of Photons and Atoms.
Identical Particles}
                                                                                
When Dirac commented that ``A photon 
interferes only with itself''\cite{Dirac1958},
he referred to the first order coherence property of the photons.
Subsequent intensity interferometry experiments\cite{Brown1954}
had revealed that photons do interfere with one another,
which are the higher order coherence properties of photons.
Such first and higher order coherence properties were also observed
for atoms in the atom interferometry experiments\cite{Berman1997}.
                                                                                
In the current ontology, the first order coherence of the atoms and
photons reveals their underlying wave and modal nature, whereas the
higher order correlation is a manifestation of the finite-Q nature
of the universe resonant cavity, resulting in ``non-pure'' spatial modes
which entangles the different field resonances.
Due to this entanglement (as reflected in the Bose-Einstein or
Fermi-Dirac statistics, for example), after emission
a photon has a tendency to merge back to the universal ``soup''
of the background photon flux during propagation, unless the photon
flux is so low that it can be described as spatially and temporally
separated monophotonic states, in which
case its degree of second order coherence $g^2(0) <1$ as is appropriate
for photons in the non-classical photon number states.
The analytical expressions for the degree of second order coherence
for bosons and fermions show different expressions according to
their respective wavefunction symmetries\cite{Scully1997},
and these statistics are only meaningful when the particle
flux is high enough.

\subsection{Resolution of the ``Schrodinger's Cat Paradox''}
                                                                                
Under the new ontology there is no longer a dichotomy between
the classical and the quantum world.  The classical systems consist
of subunits where ``wavefunction collapse'' have already been induced
by nature, through naturally occurring boundary conditions.
After a spontaneous phase transition, the overall system will be in a
quasi-stationary state and thus is stable, though its constituent parts 
may still evolve in a harmonic fashion as shown before for the
energy eigenstate.  A macroscopic object in general does not possess 
an overall quantum mechanical wavefunction that freely evolves.
                                                                                
The spontaneous nature of the phase transitions in natural systems
helps to resolve ``Schrodinger's Cat'' type of paradoxes since a
naturally occurring ``quantum measurement'' does not have to
involve a conscious observer.
The cat in question was already in a definitive Live
or Dead state before the observer openned the box, and not in a
linear superposition state of the kind: $a \cdot {\rm Live} + b \cdot
{\rm Dead}$.  The phase transition view also explains the stability
and reproducibility of these natural orders, i.e., the result
of the non-equilibrium phase transitions is insensitive to the
{\em details} of the initial-boundary conditions, and depends
only on the gross nature of these conditions\cite{Prigogine1977, Zhang1998}.
                                                                                
\section{The Hierarchical Ordering of Quantum and Classical Mechanics}
                                                                                
\subsection{Origin of Physical Laws}
                                                                                
The idea that matter throughtout the universe collectively determines 
the local inertial frames is known as Mach's principle.
A generalized version of Mach's principle demand all physical
interactions to be relational.  This characteristic is reflected in 
Feynman's formulation of a significant
fraction of dynamical laws (both quantum and classical) 
as sum of the classical action integral
over all possible paths and histories\cite{Feynman1965} i.e.,
\begin{equation}
<x_N,t_N|x_1, t_1> = \int_{x_1}^{t_N}
{\cal D} [x(t)] \exp [ i \int_{t_1}^{t_N} dt  {{L_{classical} (x, \dot{x})}
\over {\hbar}}]
.
\label{eq:12}
\end{equation}
The above space-time formulation explicitly demonstrated
that the quantum mechanical amplitude at any given location and time
is the sum of all past influences of all the amplitudes distributed
throughout space.  Classical trajectory and causal influence are
realized only because the influence of the rest of the paths
sum over to zero due to the rapid phase fluctuations.
                                                                                
Other evidence that physical laws, both classical and quantum,
are globally and resonantly selected include the fact that most laws
are deriveable from the least action or variational
principles\cite{Goldstein1980}.  This otherwise mysterious characteristic
can now be understood as that for every physical process,
the energy content of concern is always distributed globally, and the
process samples the environmental/boundary conditions of the entire space
of relevance to filter out the surviving resonant component.
                                                                                
\subsection{Symmetry and Conservation Relations}
                                                                                
The well-known relation between the symmetry of a dynamical system
and the corresponding conservation law that holds for such a system
is the celebrated Noether's theorem\cite{Noether1918,Goldstein1980}.
The existence of Noether's theorem
is another reflection that physical laws and the
matter contents have mutual dependence.
This mutual dependence is ingrained in the form of dynamical equations
(which is the reason Noether's theorem can be proven by using these
equations) since the laws/equations and the matter distribution 
are co-selected out of the the universe resonant cavity.
                                                                                
For the large-scale distribution of matter in the universe, we have
the approximate time invariance (which leads to energy conservation)
and isotropy (which leads to momentum and angular momentum conservation).
However, if we look more closely, both symmetry and conservation
on the large scale are indeed only approximate. 
                                                                                
First of all, the expansion of the universe violates the time invariance
of the matter distribution.  This could have several consequences.
If the values of the fundamental constants are determined by the
characteristics of the universe resonant cavity, depending on the
fashion of this determination the expansion could lead to the variation
of the values of these ``constants'' with time, though the case is not
settled yet of whether we have actually observed any such changes.
Another consequence is that over the long time span of the cosmic time
energy conservation is no longer guaranteed, since the matter distribution
changes with time as a result of the expansion of the universe.
This could very well be the origin of dark energy and the accelerated
expansion of the universe\cite{Reiss1998,Perlmutter1999}.
                                                                                
Secondly, as revealed by the work of Lee and
Yang\cite{Lee1956}, as well as Wu\cite{Wu1957},
in weak interactions parity conservation is violated, i.e., the laws
of physics has a preferance for ``handedness''.  If there is
indeed the interrelation between laws and matter distribution, 
this handedness in the laws
reveals that large-scale matter distribution in the
universe has a helical component in it.  This would be natural to expect
for a matter distribution originating from the Big Bang or other
initial condition for the cosmos.
The fact that the parity nonconservation only manifests in weak interactions
perhaps has to do with the fact that weak interaction is the
shortest in range of all the fundamental forces, thus is less sensitive
to the asymmetries in its immediate environment, so other longer range
influences which reflect the asymmetries of the universe is more
obviously manifest.  Furthermore, the proven conservation of 
CPT\cite{Pauli1955},
and the demonstration of CP violation 
in certain sub-class of weak interactions
shows that time-reversal symmetry is violated in these interactions as
well\cite{Lee1956}.  Therefore, 
even at the microscopic level we have the evidence of time's arrow manifest,
which is another indication that irreversible evolution and phase transitions
had played a role in the formation and selection of microscopic laws.
                                                                                
\subsection{Structures of S-Matrix Theory and Quantum Field Theories}
                                                                                
S-matrix theory was invented to circumvent certain problems 
of quantum field theories\cite{Chew1961}.
The motivation comes partly from the observation that in many scattering
experiments light quanta come in and go off as approximate plane waves.
In this theory the dynamics was not specified by a detailed
model of interaction in spacetime, but was determined by the
singularity structure of the scattering amplitudes, subject to
the requirement of maximal analyticity.  The success of the
S-matrix approach is likely to be due to the fact that the
underlying physics obeys global, modal characteristics.
The result of the calculations is expressed through particles
on the mass shell, which is equated in our ontology to the outcome
of phase transitions.  Other features of the applications of
S-matrix, such as boot-strapping and ``nuclear democracy'' in
hadron theory also reveal a distinctive modal characteristic\cite{Cao1998}.
                                                                                
Quantum field theories describe local interactions between
particles and fields.  However, certain global elements are
implicit in its formulation for processes such as scattering,
just as the S-matrix theories.  Some of the common practices in 
quantum field calculations, such as Feynman's diagrammatic approach
are integral representations of the entire phase transition process, 
described in terms of input and output states only, and omitting 
(or ``integrating out'') any detailed description of the ``on-location'' 
behavior of the interaction and particle creation/annihilation.
                                                                                
The need for renormalization (or for the manual incorporation
of the experimentally observed values of parameters into perturbative
quantum field theories to cancel certain infinities) is a
reflection of the insufficiency and non-self-consistency of
the quantum field theories.  The cause of this is partly because
the theory is local (and the renormalization allows the incorporation of
certain environmental screening effect), and partly because
it attempts to synthesize quantum theory with special relativity
(spontaneous quantum measurement phase transitions implicit
in the phenomena that quantum field theories attempt to describe
necessarily violate Lorentz invariance required by special relativity,
see further the discussion in the next section).  Empirical input 
is thus needed to construct a self-consistent field theory due to
the singular nature of phase transitions whose details
cannot be modeled in a top-down deductive type of analysis.
                                                                                
Gauge field programme emerges within the framework of quantum
field program and incorporates certain global features of field
theories.  Fundamental interactions can now be characterized by
gauge potentials and the phase of a wave function becomes
the new local variable.  In this formulation local gauge invariance
is used to derive the forms of dynamical interactions, and just as
the variational procedure for the derivation of physical laws
the success of the gauge invariance procedure reflects th resonant origin
of the relevant laws..
Spontaneous symmetry breaking was developed as a mechanism to
preserve gauge invariance when dealing with massive gauge quanta.
                                                                                
Since the late 1970s theorists gradually realized that the high-energy
effects in gauge theories can be calculated without taking the cutoff
in the normal renormalization procedure to infinity.  The cutoffs in these
Effective Field Theories\cite{Weinberg1980} serve as
boundaries separating energy regions which are separately
describable by different sets of parameters
with different symmetries.  This practice is also related to the so-called
renormalization-group appoach which explicitly deals with
the scale-dependence of certain fundamental physical interactions.
As commented by Cao\cite{Cao1999}, most of the physical theories we know
are scale-independent within its range of validity, and the scale-dependence
of parameters in field theories indicates the screening effect of the
environment, as first suggested by Dirac, and also indicates the
smoothness and homogeneity of the variation of these interactions with scale.
                                                                                
\subsection{Issues about Covariance and the Absolute Reference Frame}
                                                                                
We expect that the establishment of fundamental laws and
the selection of fundamental constants through the resonant interaction
in the universe according to Mach's principle are
processes which are not constrained by the speed of light limit:
For otherwise the mere size of the universe and the cycle of
propagations needed to resonantly establish a fundamental constant
would prevent these constants and laws to be universal.
Similarly, the variational approach for the selection of physical laws
only works if all possible paths are explored simultaneously/instantaneously.
                                                                                
The speed-of-light limit appears to hold strictly only for classical
processes.  In classical physics, the influences of the nonclassical paths,
which involve superluminal signal propagation, cancel each other
out\cite{Feynman1965}, so the superluminal effect is not apparent.
                                                                                
For quantum physics, both the measurement and the
non-measurement type, the effect of space-time paths requiring
superluminal signal propagation can no longer be ignored.
The quantum measurement processes, especially the
Einstein-Podolsky-Rosen type or delayed-choice\cite{Scully1997}
type of experiments, showed both the global extent of the wavefunction
and the superluminal nature of the wavefunction collapse.
That this is also true during most of the unitary evolution stages
of a quantum state is less obvious, but is reflected nonetheless in
the quantum mechanical rules of state evolution (equations \ref{eq:5},
\ref{eq:6}), in Feynman's path integral formulation of quantum
mechanics (equation \ref{eq:12}, which contains the integration over
the classical paths at subluminal speed and over nonclassical paths
at superluminal speed), as well as in Feynman diagrams
for most of the virtual paths.
                                                                                
Covariance in relativistic quantum mechanics
appears to hold only for the {\em results} of quantum measurements
(local events), and not for the evolution of states themselves
(global events in general)\cite{Aharonov1981}, 
even though many of the formulations
of QED are supposed to be ``manifestly covariant'' (such as Feynman's
path integral approach, which as we have just commented still
contains the contributions from superluminal pathways).
Especially for states which have not completed the phase transition,
such as many quantum tunneling processes and certain optical
evanescent-wave propagation, the speed-of-light limit is often found to be
violated\cite{Steinberg1993,Enders1993}, and these
processes appear to occur ``on borrowed energy''
from their environment.
                                                                                
Further more, there appears to be the need for a universal reference
frame in order for the phase transitions to happen with respect to.
The Aharonov-Bohm effect\cite{Aharonov1959} which shows the influence to 
the phase of the wavefunction in regions where the nominal field strength 
is absent demonstrates the substantiality of a global potential function.
In fact, the reference frame established by the entire content of the
universe is a convenient one, if the universe is finite.  This reference
frame obviously possesses the relational nature demanded by Mach's principle.
                                                                                
\section{Conclusions}
                                                                                
The new ontological view on the quantum measurement problem based
on the generalized Mach's principle and on nonequilibrium phase transitions,
is shown to lead to coherent interpretations of a wide spectrum of
well-known phenomena in both the quantum and the classical world.
Such global phase transitions are likely to be responsible for
the structure and forms of physical laws, the values of fundamental
constants, as well as the hierarchies of phenomena we observe
in the microscopic and macroscopic world.  If proven correct,
the insight thus gained from this picture could provide intuitive
guidance to the efforts of the attempted unification of fundamental
interactions, as well as to the resolution of issues at the interface of
physics and cosmology such as the accelerated expansion of the universe
and th origin of dark energy..
                                                                                
\smallskip 

This research was supported in part by funding from the Office of
Naval Research.

\end{document}